\documentclass[aps,showpacs,amsfonts,amsmath, amssymb,twocolumn,floatfix,superscriptaddress]{revtex4-1}
\usepackage[final]{graphicx}
\usepackage{color}
\usepackage{ulem}

\newcommand{\be}{\begin{equation}}
\newcommand{\ee}{\end{equation}}
\allowdisplaybreaks

\DeclareSymbolFont{bbgreek}{U}{bbold}{m}{n}
\DeclareMathSymbol{\bbmu}{\mathbb}{bbgreek}{'26}
\DeclareMathSymbol{\bbeps}{\mathbb}{bbgreek}{'17}
\begin{document}

\title{Interplay of   roughness/modulation  and curvature at proximity}

\date{\today}

\author{Matthias Kr\"uger}
\altaffiliation{Present address: 4th Institute for Theoretical Physics, Universit\"at Stuttgart, Germany and Max Planck Institute for Intelligent Systems, 70569 Stuttgart, Germany
}
\affiliation{Massachusetts Institute of Technology, Department of
  Physics, Cambridge, Massachusetts 02139, USA}
\author{Vladyslav~A. Golyk}
\affiliation{Massachusetts Institute of Technology, Department of
  Physics, Cambridge, Massachusetts 02139, USA}

\author{Giuseppe Bimonte}
\affiliation{Dipartimento di Scienze Fisiche, Universit{\`a} di Napoli Federico II, Complesso Universitario MSA, Via Cintia, I-80126 Napoli, Italy and INFN Sezione di Napoli, I-80126 Napoli, Italy}

\author{Mehran Kardar}
\affiliation{Massachusetts Institute of Technology, Department of
 Physics, Cambridge, Massachusetts 02139, USA}

\begin{abstract} 
We show  that roughness or  surface modulations change the distance dependence of (power-law) interactions between curved objects at proximity. The modified scaling law is then simply related to the order of the first non-vanishing coefficient of the Taylor expansion of the distribution of separations between the surfaces.  The latter can in principle be estimated by scanning measurements, or computed for well characterized modulations, and then used to predict short-distance scaling behavior in disparate experiments. For example, we predict that the radiative heat transfer between a rough sphere and a plate   approaches a constant with decreasing separation.  Similar saturation is expected for the Casimir force between dielectric or metallic surfaces with appropriate modulations over distinct length scales.
\end{abstract}

\bibliographystyle{plain}

\maketitle

The Proximity Approximation (PA) has long served as a useful guide for estimating interactions between closely spaced objects with curved surfaces.   
Originally introduced  by Derjaguin~\cite{Derjaguin34} to compute van der Waals forces between colloidal particles, the PA relates  the interaction between curved objects at close separations to the corresponding interaction between two flat surfaces, over an area determined by the local radii of curvature. The PA has been successfully used in experimental studies in a variety of fields, such as quantum Casimir forces~\cite{Lamoreaux97,Mohideen98,Bordag}, classical Casimir forces in a fluid near a critical point~\cite{Hertlein08}, near field radiative heat transfer~\cite{Sheng09,Rousseau09}, and  interactions among nuclei~\cite{Blocki77}.  Recently it has also been applied to the gravitational force in searches of non-Newtonian gravity~\cite{Decca09}.
 
In this paper, we utilize PA to investigate scaling of the interaction between two surfaces whose local radii of curvature can be decomposed as the sum of components which vary along the surface on well separated length scales. We note that this situation encompasses the experimentally relevant cases of two large rough objects, such that the correlation length of the roughness is much smaller than the characteristic radius of curvature of the surface,  or alternatively the case of a surface with large average radius of curvature modulated by small structures fabricated by the experimenter. We find that at proximity a subtle interplay  between roughness/modulation and  the  global curvature of the surfaces  leads to a drastic change in the distance dependence of the interaction,  compared to that for perfectly smooth (structureless) surfaces.  
The modified scaling law is found to depend in a simple way on the order of the first non-vanishing term in the Taylor expansion of the height distribution function of the surfaces. The latter is a geometric feature of  the surface that can be either computed for well characterized modulations, or experimentally obtained via scanning probes of rough surfaces,
and then used  to predict the short distance scaling of results in a multitude of experiments.
We discuss  the example of radiative heat transfer, and  show how   modulations (of various shapes) or roughness limit  the maximum amount of non-contact heat transfer that can be achieved between two surfaces in close proximity.  Applications to  the quantum Casimir force shall also be discussed briefly.

\begin{figure}
\includegraphics[width=0.9\linewidth]{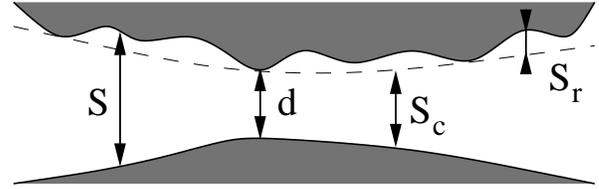}
\caption{\label{fig:0} Profiles of two gently curved surfaces: The local separation $S$ is the sum
of a slowly varying component $S_c$ due to overall shape, and a rapid part due to roughness/modulation; $d$ is the distance of closest approach.
}
\end{figure}

Consider two objects with surface-to-surface separation $S=S(x,y)$, where $(x,y)$ are Cartesian coordinates in a plane that separates the two surfaces, as in Fig.~\ref{fig:0}.   The unspecified  interaction $I$ between them may depend in a complicated way on the geometry of the objects. However, we assume that at close proximity it is dominated by the local properties of the surfaces near the point of closest approach. In the proximity approximation $I_{PA}$ the net interaction is given by 
\begin{equation}
I_{PA}=\int dx\int dy I_{pp}\left(S(x,y)\right),\label{eq:PTA}
\end{equation}
where $I_{pp}(S)$ denotes the interaction between two parallel plates at separation $S$ (per surface area). 
It is reasonable to expect  that Eq.~\eqref{eq:PTA} becomes exact for nearly flat surfaces, i.e. in the limit $d/\rho \rightarrow 0$, where $\rho$ sets the scale for the local radii of curvatures. 
Indeed, in certain cases Eq.~\eqref{eq:PTA} represents the leading term of an expansion of the interaction in powers of the gradient of $S$~\cite{Fosco11, Bimonte12,Bimonte12b, Golyk12} (see Eq.~\eqref{eq:grad} below). This implies that  Eq.~\eqref{eq:PTA} is indeed asymptotically exact in the limit  $d/\rho\rightarrow 0$, the leading
curvature correction being of higher order in $d/\rho$. Equation~\eqref{eq:PTA} can be alternatively recast in terms of the {\it height distribution function}~\cite{Decca05,Broer12} $f(s)\equiv \int dx dy\, \delta(S(x,y)-d-s)$, as
\begin{equation}
I_{PA}=\int ds f(s-d) I_{pp}(s)\,.\label{eq:PTA2}
\end{equation}
Note that $f(s)$ is measured with respect to the distance $d$ of closest approach, such that $f(s)=0$ for $s<0$. It is thus a function of shape (and orientation) of the objects, and independent of their separation.
Equation~\eqref{eq:PTA2} highlights the basic assumption of PA in neglecting the detailed topology of the surface (e.g. gradients), being sensitive only to the projected surface area at separation $s$. 
While in general only an approximation, Eq.~\eqref{eq:PTA2} can be used to understand the qualitative effects of modulation or roughness in concrete experimental setups;
the function $f(s)$ can be easily extracted from atomic force microscope (AFM) scans of the surfaces~\cite{Decca05}, or  computed analytically for model surfaces (e.g., below).

In what follows, we assume that $I_{pp}(d)$ diverges with a simple power-law for $d\to0$, 
\begin{equation}
I_{pp}(d)=\frac{\alpha}{d^{\nu}}.\label{eq:kernel}
\end{equation} 
For example, $\nu=3$ for the quantum Casimir energy between perfectly conducting surfaces~\cite{Bordag}, while $\nu=2$ for the critical Casimir energy involving binary fluids~\cite{Hertlein08} or radiative heat transfer~\cite{Volokitin01}. The dimensional coefficient $\alpha$ depends on the specific system under study.

Starting from Eq.~\eqref{eq:kernel}, it is easy to verify that the asymptotic behavior of $I_{PA}$ at close proximity is determined by the first non-vanishing coefficient $f^{(n)}(0)$ of the Taylor expansion  of  $f$ near $s=0$, as summarized in Table~\ref{table:1}.
Considering for example a sphere of radius $R$ in front a plate, we have 
\begin{equation}
f_{S}(s)=2\pi(R-s)\,,\label{eq:fSp}
\end{equation} 
for $0\leq s \leq R$ and zero otherwise; $f_{S}(s=0)$ is finite (case $\mathcal{C}=1$ in Table~\ref{table:1}), leading to well known results: $I_{PA}\propto d^{-2}$ for $\nu=3$, and $I_{PA}\propto d^{-1}$ for $\nu=2$. Another example is a square base pyramid of height $h$ and base length $l$, with axis perpendicular to a plate, for which  
\begin{equation}
f_{P}(s)=2s\frac{l^2}{h^2},\label{eq:fp}
\end{equation}
for $0\leq s\leq h$ and zero else.
 The sharp tip of the pyramid has vanishing surface density (case 2), such that PA yields a distinctly different behavior than for the sphere-plate configuration: A $1/d$-divergence for $\nu=3$, or a logarithmic one for $\nu=2$. 
 Although the PA is not asymptotically exact for singular surfaces with sharp tips,  
the above conclusion agrees qualitatively with exact computations of the Casimir force and heat transfer in the cone-plate geometry~\cite{Maghrebi11c,McCauley}.
\begin{table}
\begin{ruledtabular}
\begin{tabular}{|c|c|l|c|}
\hline
Case&$\lim_{d\to0}I_{PA}=$&If&Examples\\
\hline\hline
${\cal{C}}=1$&
\begin{math}
\begin{array}{cc}
\mathcal{O}(d^0)&\nu<1\\
-\alpha f(0)\log\left[\frac{d}{d_0}\right]&\nu=1\\
\frac{\alpha f(0)}{\nu-1}\frac{1}{d^{\nu-1}}&\nu>1\\
\end{array}
\end{math}
&$f(0)>0$&\includegraphics[height=0.5cm]{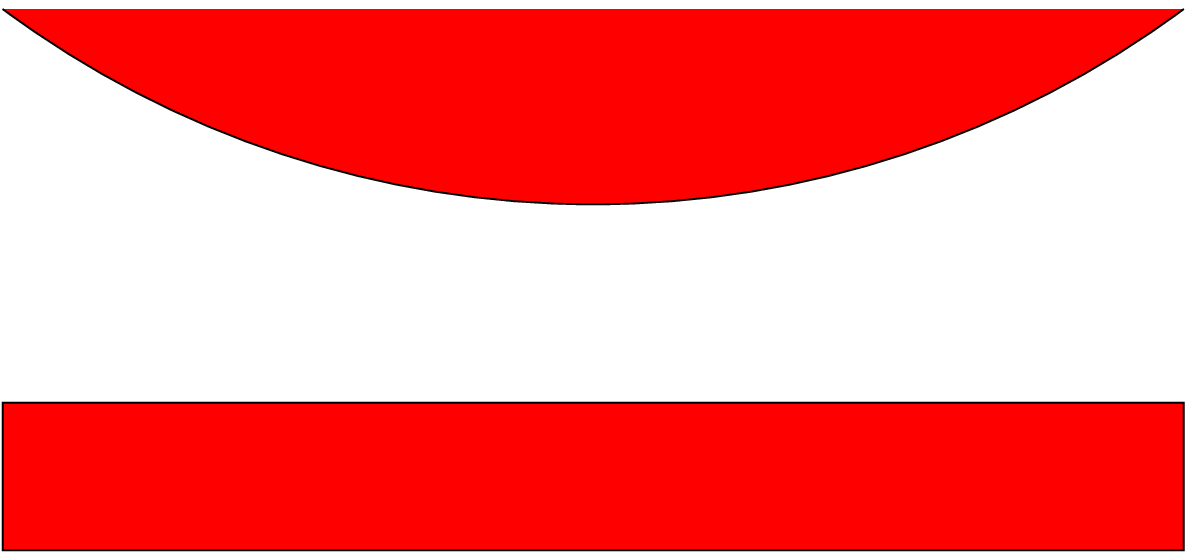}\\
\hline
${\cal{C}}=2$&

\begin{math}
\begin{array}{cc}
\mathcal{O}(d^0)&\nu<2\\
-\alpha f'(0)\log\left[\frac{d}{d_0}\right]&\nu=2\\
\frac{\alpha f'(0)}{\nu-2}\frac{1}{d^{\nu-2}}&\nu>2\\
\end{array}
\end{math}
&$f(0)=0$&
\begin{tabular}{c}\includegraphics[height=0.5cm]{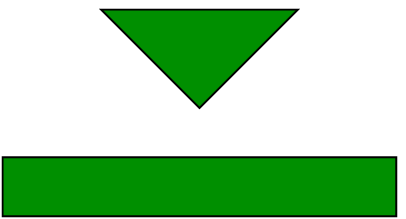}\\ \includegraphics[height=0.5cm]{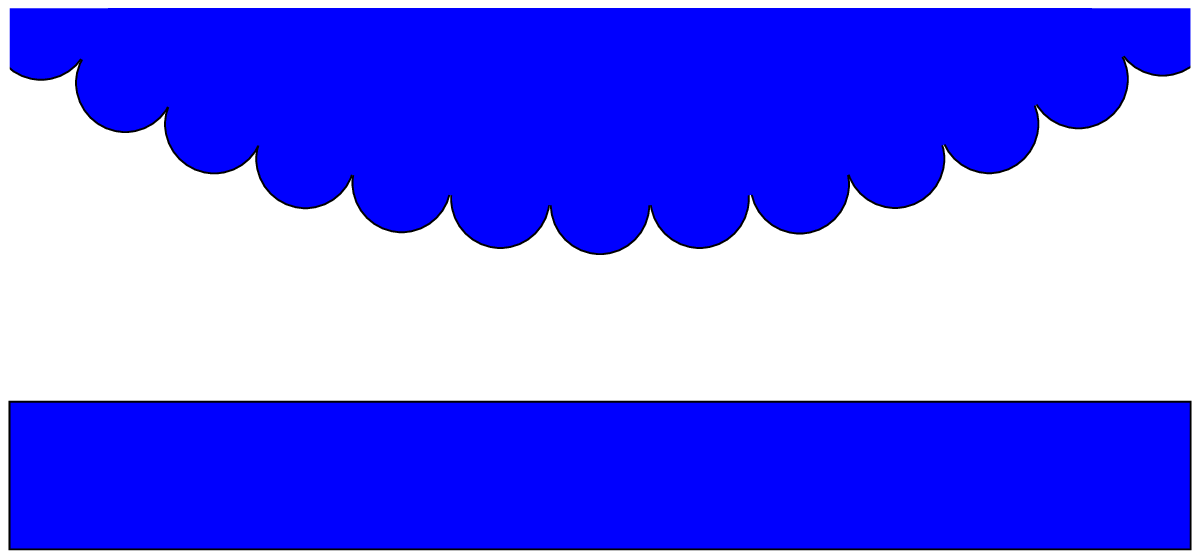}\end{tabular}
\\
\hline
${\cal{C}}=3$&
\begin{math}
\begin{array}{cc}
\mathcal{O}(d^0)&\nu<3\\
-\frac{\alpha}{2} f''(0)\log\left[\frac{d}{d_0}\right]&\nu=3\\
\frac{\alpha f''(0)}{2(\nu-3)}\frac{1}{d^{\nu-3}}&\nu>3\\
\end{array}
\end{math}
&
\begin{math}
\begin{array}{c}f(0)=\\f'(0)=0 \end{array}
\end{math}
&\includegraphics[height=0.5cm]{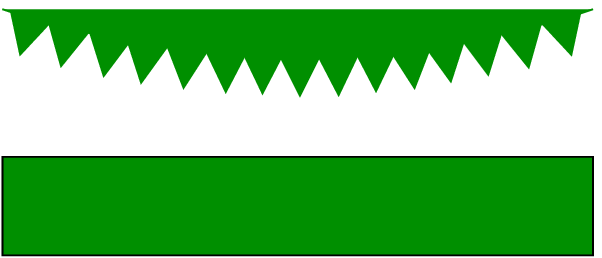}\\
\hline
$\vdots$&&&\\
\hline
${\cal{C}}=n$&
\begin{math}
\begin{array}{cc}
\mathcal{O}(d^0)&\nu<n\\
-\frac{\alpha f^{(n-1)}(0)}{(n-1)!}\log\left[\frac{d}{d_0}\right]&\nu=n\\
\frac{\alpha f^{(n-1)}(0)}{(n-1)!(\nu-n)}\frac{1}{d^{\nu-n}}&\nu>n\\
\end{array}
\end{math}
&
\begin{math}
\begin{array}{c}f(0)=\\\dots=\\f^{(n-2)}=0 \end{array}
\end{math}
&\\

\hline
\end{tabular}
\end{ruledtabular}
\caption{Asymptotic behavior of the interaction between two surfaces, as determined by the order of the first non-vanishing Taylor coefficient  $f^{(n)}(0)$ of the height distribution function of the surfaces. 
The interaction scales as  $I_{pp}(d)=\alpha/d^{\nu}$ for two parallel surfaces. The length $d_0$ depends on higher derivatives of $f$.}
\label{table:1}
\end{table}

We now turn to the experimentally relevant case of  rough or modulated  surfaces.  The surface-to-surface separation $S$ can be decomposed as $S=S_c+S_r$, where $S_c$ is the separation between two smoothened surfaces with general radius of curvature ${\cal R}$, and $S_r$ is the additional shift resulting from  roughness or modulation (see Fig.~\ref{fig:0}). If the length scale  $\xi$  characterizing the roughness or the modulation  is much smaller than the characteristic size  $L$ of the interaction area (typically of order $L \sim \sqrt{d {\cal R}}$), i.e., $\xi \ll L\ll {\cal R}$, we find that  the height distribution function $f(s)$ for $S(x,y)$ is well approximated by the {\it convolution} of the corresponding functions  $f_{c}$ of the smoothened surface, and $f_{r}$ (normalized by a unit surface area) for  roughness/modulation, i.e.
\begin{equation}
f(s)=\int_{0}^s ds' f_c(s')\,f_r(s-s').\label{eq:convolution}
\end{equation}
This result also obtains from the commonly-used two-step application of the PA~\cite{Bordag}, in which the PA is first used to estimate the correction arising from roughness or modulation (per unit area) to the interaction between parallel plates, and then used again to estimate the effect of curvature. 
It has been noted (see e.g. Ref.~\cite{Broer12}) that if the experimental conditions are such that the approximate formula of Eq.~\eqref{eq:convolution} is not valid for the surfaces of interest,   measurements might not be fully repeatable, as they would be sensitive to e.g. uncontrollable relative displacements of the two surfaces in the $(x,y)$ plane.

The asymptotic scalings in Table~\ref{table:1} also apply to the combined distribution function in Eq.~\eqref{eq:convolution}, where we note an intriguing relation. Restricting to regular, analytic functions near $s=0$, the case number of the combined function $f$ in Eq.~\eqref{eq:convolution} is precisely the sum of the case numbers of the individual functions $f_c$ and $f_r$,
\begin{equation}
\mathcal{C}(f)=\mathcal{C}(f_c)+\mathcal{C}(f_r).\label{eq:CN}
\end{equation}
Equation~\eqref{eq:CN} can be derived with the convolution theorem for Laplace-transforms, and combined with Table~\ref{table:1} provides a means for obtaining the modified scalings. 

\begin{figure}
\includegraphics[width=0.9\linewidth]{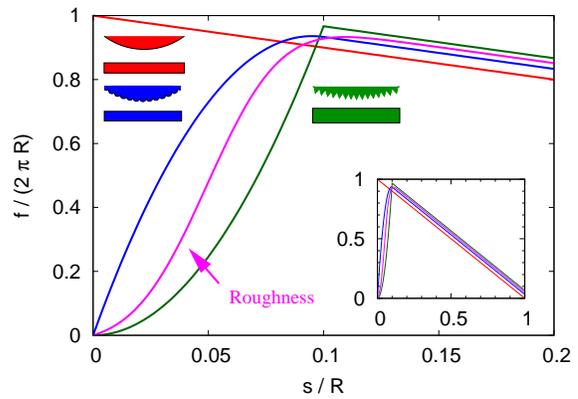}
\caption{\label{fig:1}Height distribution functions for model surfaces (indicated by icons/text) in the sphere-plate geometry; for modulated surfaces $h=R/10$, while $\sigma=R/40$ and $s_0=2\sigma$ for the rough surface. The larger range of $s$ in the inset emphasizes that the  impact of deformations is mostly for small $s$.}
\end{figure}

Figure~\ref{fig:1} shows $f(s)$ for the case of a sphere with different types of modulation in front a plate \footnote{While for simplicity we consider a flat plate, both plates could be modulated as the results only depend on the relative separation.}, where the red curve  is for the smooth sphere of Eq.~\eqref{eq:fSp}. If the surface of the sphere is covered with (square base) domes of height $h$ ($f_r \equiv f_{D}=2 (h-s)/h^2$), $f$ is found analytically with the help of Eqs.~\eqref{eq:fSp} and \eqref{eq:convolution}, as 
\begin{equation}
f_{SD}(s\leq h)=\frac{2 \pi  s \left(6 h R-3 h s-3 R s+s^2\right)}{3 h^2}\label{eq:f0s}.
\end{equation}
As we see, $f_{SD}(0)$ vanishes, however its first derivative $f_{SD}'(0)$ does not, see the blue curve in Fig.~\ref{fig:1}. Therefore, the dome-like modulation ($\mathcal{C}=1$), yields $\mathcal{C}=2$ when convoluted with the sphere-plate geometry. 

Suppose now that the sphere is covered with pyramids of height $h$ (again square based for ease of tiling, as in Eq.~\eqref{eq:fp}).  After convolution with the sphere we now find
\begin{equation}
f_{SP}(s\leq h)=\frac{2 \pi  s^2\left(3 R -s\right)}{3 h^2}\,,\label{eq:SpPy}
\end{equation}
which is order $s^2$ (green curve in Fig.~\ref{fig:1}). Sharp tips ($\mathcal{C}=2$), give $\mathcal{C}=3$ when  convoluted with a curved surface. 

We now consider surface roughness, which in general depends on materials and manufacturing~\cite{Aspnes79,Benardos03,Persson05}. 
While small deformation of amplitude $w \ll d$ can be treated perturbatively~\cite{Neto05,Biehs10}, 
perturbative methods fail on close approach when $d$ becomes comparable to $w$,
and the PA provides a better route~\footnote {Reference~\cite{Broer12} shows that estimates of the Casimir force obtained by means of the PA  are in good agreement  with experiments on rough surfaces for  $d \sim w$.}.  For practical ease, the height distribution function $f_r$ of a rough surface will be approximated by a Gaussian~\cite{Broer12},
\begin{equation}
f_{R}(s\geq 0)=\frac{1}{\mathcal{N}\sigma\sqrt{2\pi}}\exp\left[-\frac{\left({s-s_0}\right)^2}{2\sigma^2}\right]
\label{eq:RS},
\end{equation}
where $\sigma$ is the standard deviation and $\cal N$ is chosen such that $\int_0^\infty ds f_R(s)=1$. 
The distribution $f_r$ has no sharp boundaries (as the peaks and grooves of the surface can in principle have arbitrary extension), and we introduce an additional parameter $s_0$ to set $s=0$, i.e., the ``touching distance.'' This parameter sets the height of the highest peak, which is touched first on approach and acts as a shift on the separation axis. 
The truncated Gaussian in Eq.~\eqref{eq:RS} corresponds to case 1, although with a possibly small value of $f_R(0)$. After convolution with the curvature of the sphere, we find that the height distribution function of  the rough sphere is linear near $s=0$ ($\mathcal{C}= 2$). The magenta curve in Fig.~\ref{fig:1} starts linearly, but with a small slope (for the chosen value of $s_0=2\sigma$), such that the rough surface is intermediate  between the dome-like and pyramid-like modulations considered earlier.

\begin{figure}
\includegraphics[width=0.9\linewidth]{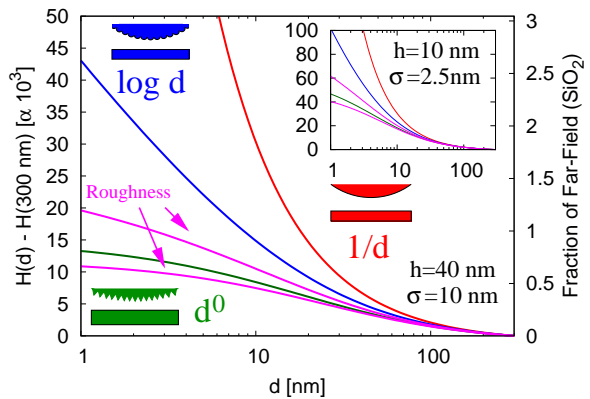}
\caption{\label{fig:2}Heat transfer between a flat plate and deformed model spheres  (as shown by icons/text) of radius R=50$\mu$m. The amplitude $h$ for modulations (dome or pyramid),
and the width $\sigma$ for roughness are as indicated, and 4 times smaller in the inset.
The two curves for roughness correspond to $s_0=2\sigma$ (upper) and $s_0=3\sigma$ (lower).}
\end{figure}

As a specific example,  we  consider  radiative heat transfer $H$~\cite{Polder71} between a rough or modulated sphere and a (perfectly flat) plate, demonstrating that  roughness or modulation  limit the maximum achievable non-contact energy transfer,  an issue of much current technological and experimental interest~\cite{Sheng09,Rousseau09,Ottens11}.  We
note that for heat transfer, the PA has been shown to approach the exact result at small separation for a sphere in front a plate~\cite{Kruger11, Otey}, and for two spheres~\cite{Sasihithlu11}.  
As shown in Ref.~\cite{Golyk12}, an asymptotic expansion in powers of the gradient of $S$ exists for heat transfer $H$, which for small separations leads to  
\begin{equation}
H= H_{PA}+\int ds \,\,\beta \,\,g(s-d)  H_{pp}(s) +\cdots\, .\label{eq:grad}
\end{equation}
The second term in Eq.~\eqref{eq:grad} represents the leading curvature correction to PA, with $g(s)\equiv \int dx dy\, \delta(S(x,y)-d-s)\nabla S \cdot \nabla S$ encoding the mean squared gradient at $s$. The dimensionless coefficient $\beta$ 
depends on  material and temperature, and is typically of order unity~\cite{Golyk12}. The function $g(s)$ for a rough or modulated surface can be found  in a manner similar to Eq.~\eqref{eq:convolution}, and then Eq.~\eqref{eq:grad} allows us to judge whether the scaling laws obtained from the PA are asymptotically exact for specific shapes.

For small separations $d$, the radiative transfer between  parallel plates (per unit area)
 diverges as a power law with exponent $\nu=2$~\footnote{After subtraction of the far field contribution to $H$, for SiO$_2$ at room temperature, this scaling law extends up to $d\alt300$~nm~\cite{Golyk12}}.
Figure~\ref{fig:2} illustrates the transfer as given by PA for the cases shown in Fig.~\ref{fig:1}. In all cases we subtracted from $H$ the far field contribution (computed at the reference separation $d=300$ nm).
The smooth sphere shows the expected divergence of $1/d$ (red curve); dome-like modulations (with  $f(s)$ in Eq.~\eqref{eq:f0s}) reduce this divergence into a logarithmic form (blue curve), such that 
\begin{equation}
\lim\limits_{d\to0} H_{SD}(d)=-4 \pi \alpha \frac{R}{h} \log\frac{d}{d_0}.\label{eq:SDt}
\end{equation}
For this type of modulation, the gradient correction in Eq.~\eqref{eq:grad} is of order $d^0$,  and therefore  Eq.~\eqref{eq:SDt} becomes exact for vanishing separation.
 
\begin{figure}[ttt]
\includegraphics[width=0.6\linewidth]{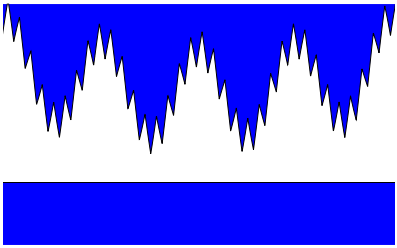}
\caption{\label{fig:4} 
Example of a surface pattern (not to scale) with $\mathcal{C}=4$. For such a geometry, the quantum Casimir force for real materials (and in general any interaction with $\nu=3$) is predicted to level off to a constant value at small separations.
}
\end{figure}

A surface with pyramid-like modulations ($f(s)$ as per Eq.~\eqref{eq:SpPy}) shows an even more drastic slow down: according to Table~\ref{table:1}, the transfer levels off to a constant value (green curve).  However,  evaluation of the gradient term in Eq.~\eqref{eq:grad}  shows that this correction is now of the same order as the PA term. From this we infer that  PA is not exact in this regime, while the predicted saturation to a constant is presumably correct.
{\it For rough surfaces} the height distribution function is somehow intermediate between those for dome-like and pyramid-like modulations. Confirming this expectation,  we find that $H$ diverges logarithmically for rough surfaces, but only with a small prefactor. Depending on $\sigma$ and $s_0$ in  Eq.~\eqref{eq:RS}, this divergence may not be visible in practical cases, giving the appearance
of saturation as in the case of  pyramid-like modulations. 
Reducing the amplitude of roughness/modulation does not change the asymptotic behavior, but
differences from a smooth sphere become visible only at closer separations (inset of Fig.~\ref{fig:2}).

On a practical note, the  right axis in Fig.~\ref{fig:2} shows the ratio to the classical far field ($d\to\infty$) transfer for SiO$_2$, for $T_1=0$ and $T_2=300$~K (which is $\sim 4.2\mu W$~\cite{Kruger11}). With $\alpha=0.2558$nW~\cite{Golyk12}, and taking  $R=50 \mu$m and $\sigma=10$~nm, 
we predict that (rather disappointingly) heat transfer can only  be approximately doubled by reducing the separation to $1$~nm (for the considered range of $s_0$).

In principle saturation can also be achieved for Casimir forces with $\nu\geq3$ by appropriate
surface fabrication. Figure~\ref{fig:4} depicts a combination of three height profiles, a sphere with a smooth modulation and an additional sharp modulation. If the corresponding three lengths are well separated, e.g. for $R=100 \mu$m, and modulations lengths of 1$\mu$m and 10~nm, respectively, the convolution in Eq.~\eqref{eq:convolution} (and also Eq.~\eqref{eq:CN}) can be applied two successive times, to yield ${\cal C}=1+1+2$. The Casimir force between two such dielectric or metallic objects, for which $\nu=3$~\cite{Bordag}, should saturate to a constant value.

We have shown that the interplay between roughness or modulation and curvature leads to drastic modifications of the short distance behavior of  interactions  between two surfaces. By simple computations  based on the PA, we find that the scaling law for the interaction is determined by the first non-vanishing Taylor coefficient of the height distribution function of the surfaces. If  modulations  occur over widely  distinct length scales, their effect on the predicted scaling law is additive.

We thank P.~L.~Sambegoro, G.~Chen, R.~L. Jaffe, T. Emig,
 M.~F. Maghrebi, M.~T.~H. Reid and N. Graham for helpful
discussions. This research was supported by the DFG grant No. KR 3844/2-1,
NSF Grant No. DMR-12-06323, DOE grant No. DE-FG02-
02ER45977, and the ESF Research Network CASIMIR.

%\bibliography{cas}
%\bibliographystyle{apsrev4-1}
\end{document}